\documentclass[12pt]{iopart}

\newcommand{\be}{\begin{eqnarray}}
\newcommand{\ee}{\end{eqnarray}}
\newcommand{\bea}{\begin{eqnarray}}
\newcommand{\eea}{\end{eqnarray}}

\usepackage{graphicx}
\usepackage{bm}
\usepackage{color}
\usepackage{slashed}
\usepackage{cite}
\usepackage{graphicx}
\usepackage{multicol}
\usepackage{graphicx}
\usepackage{color}
\usepackage{slashed}
\usepackage{CJKutf8}
\begin{document}
\begin{CJK}{UTF8}{<font>}
\title{Phase transition grade and microstructure of AdS black holes in massive gravity}

\author{Guan-Ru Li$^{1}$, Guo-Ping Li$^{2}$, Sen Guo$^{*1}$}

\address{$^1$Guangxi Key Laboratory for Relativistic Astrophysics, School of Physical Science and Technology, Guangxi University, Nanning 530004, China\\
         $^2$School of Physics and Astronomy, China West Normal University, Nanchong 637000, China}

\ead{r1055043727@126.com; gpliphys@yeah.net; sguophys@126.com}
\vspace{10pt}
\begin{indented}
\item[]Jun 2022
\end{indented}

\begin{abstract}
Considering that under the framework of the massive gravity theory, the interaction between the mass gravitons and Schwarzschild black hole (BH) could make it carry a scalar charge, the phase transition process caused by this scalar charge is investigated in this analysis. The phase transition grade and microstructure of those BHs are investigated from both macroscopic and microscopic points of view. From the macroscopic point of view, it is found that Ehrenfest equations are satisfied at the phase transition critical point, which implies that the phase transition grade of those BHs is second-order. Based on the BH molecules model and Landau continuous phase transition theory, the phase transition of those BHs from the microcosmic point of view is analyzed. The critical exponents obtained from the two perspectives are consistent. By investigating the Ruppeiner geometry, the microstructure feature of those BHs is revealed. These results suggest that the phase transition of BH in massive gravity is a standard second-order phase transition at the critical point, and the microscopic details of those BHs are different from the RN-AdS BH in standard Einstein gravity.
\end{abstract}


\section{Introduction}
\label{intro}
\par
Observation results of the gravitational wave of Laser-Interferometer Gravitational Wave-Observatory (LIGO) collaboration conform to the predictions of General Relativity (GR) \cite{1}. Although the success of GR is unprecedented, it is meaningful and challenging to find an effective alternative to this theory. Because there are still some thorny problems that need to be explained more reasonably to reconcile gravity, cosmology, and particle physics. Observations show that our Universe is in the stage of accelerated expansion, and the driving mechanism is not clear until now. There is an irreconcilable contradiction between GR and quantum field theory in the vacuum energy required to drive accelerated expansion. An option is to accept that the gravitational laws get changed at a cosmological distance and time scales. The gravity at the low energy stage needs to be modified. Massive gravity theory is a theory based on giving graviton mass to modify the laws of gravity at the infrared stage. Considering giving the graviton mass, Fierz and Pauli obtained a new gravity theory by modifying the Einstein field equation \cite{2}. The new gravity theory containing gravitons with mass is unique if the physical applicability is considered. By investigating massive and mass-less Yang-Mills and gravitational fields, Dam $et~al.$ and Zakharov (vDVZ) found that the linear massive gravity theory can not degenerate to the Einstein theory in the $m\rightarrow0$ limit, which is the vDVZ discontinuity, which is the vDVZ discontinuity \cite{3,4}. The discontinuity is due to the existence of a coupled scalar even under the $m\rightarrow0$ limit. The nonlinearities of the massive gravity theory were derived by Vainshtein, who proposed that a small mass graviton can ensure the massive gravity theory returns to Einstein's theory in strong nonlinear characteristics \cite{5}. This makes massive gravity theory reasonable and acceptable. By arguing some completely nonlinear massive gravity theories, Boulware $et~al.$ showed that the nonlinear features usually introduce an abnormal scalar, which is known as the Boulware-Deser ghost. It is indicated that the complete regression of massive gravity theory to GR depends on the abnormal scalar of the nonlinear part to counteract the scalar of the linear part \cite{6}.

\par
In the past two decades, the modified gravity theory based on the framework of giving graviton mass has emerged one after another including the massive gravity theory with infrared modification characteristics. It defaulted that a modified gravity theory like massive gravity theory is acceptable if it only applies below the ultraviolet cutoff. Arkani-Hamed $et~al.$ investigated massive gravity theory with the help of effective field theory and showed a method of restoring gauge invariance to massive gravity. In the form of effective field theory, there is generally a maximum ultraviolet cutoff, accompanied by graviton with mass \cite{7}. Some solutions with continuous ``non-co-diagonal'' characteristics in massive gravity theory are presented by Damour $et~al.$, founding that these solutions can correspond to the accelerated expansion of the Universe and the mass terms in this case of symmetry breaking may not come from brane configurations \cite{8}. For the general Lorentz-violating graviton mass under the three-dimensional Euclidean group, Dubovsky showed that the property of ultraviolet insensitivity can make the massive gravity theory remain well behaved even after the inclusion of arbitrary higher dimension operators \cite{9}. By contrasting the gauge theory and massive gravity theory, Hooft proposed that gravitons can obtain mass through a mechanism similar to the Brout-Englert-Higgs mechanism. The ultraviolet boundary phenomenon means that the gravitons with mass have a scalar interaction form in the infrared domain and have a mild interaction form in the far ultraviolet \cite{10}. Soon afterward, considering that coupling derivative scalar can spontaneously break the Lorentz symmetry and make the graviton have mass, Bebronne $et~al.$ showed black hole (BH) solutions in massive gravity. They found that a correction of scalar charge will exist in the Schwarzschild solution, which makes gravitational potential decay slower than $1/r$ in a certain distance range \cite{11}. The Lorentz-violating massive gravity is usually accompanied by instantaneous interaction, BHs in massive gravity have "hair". The property of BHs in massive gravity does differ from those in GR.

\par
BH thermodynamics has been extensively studied and has been booming since its birth. By investigating the Schwarzschild BH of anti-de Sitter (AdS) space-time, which is the solution of Einstein field equations with negative cosmological constant, Hawking and Page first proposed that phase transition can occur in such a thermal system \cite{13}. This phase transition is called Hawking-Page phase transition. The possible mechanism behind this phase transition in the Schwarzschild AdS BH system and related thermodynamic phenomena are discussed \cite{14,15,16,17,18,19,20,21}. Based on the research method of the extended phase space in which the cosmological constant is regarded as the thermodynamic pressure, Reissner-Nordstr\"{o}m AdS (RN-AdS) BH is found to have a phase transition similar to the classical van der Waals (vdW) fluid system. By analogy with the vdW system, Kubiz\v{n}\'{a}k $et~al.$ proposed the critical behavior and critical exponents of the RN-AdS BH system. It is found that RN-AdS BH and vdW systems are completely consistent and the results of critical exponents are independent of the number of dimensions \cite{22}. Subsequently, Spallucci $et~al.$ constructed Maxwell's equal area law in the RN-AdS system. They found that large-small BH phases exist in the RN-AdS system, corresponding to the gas-liquid two states in the vdW system respectively \cite{23}. The large number of researches focused on the BH phase transition caused by charge, including thermodynamic features of BH with different electric field sources \cite{24,25}, thermodynamic criticality showed by Ehrenfest scheme \cite{26,27}, Joule-Thomson expansion process of BH \cite{28,29}, continuous behavior of the phase transition \cite{30,27}.

\par
On the other hand, based on the microscopic fluctuation theory in classical thermodynamics, the Ruppeiner geometry method is introduced into the AdS BH system to discuss phase transition behavior \cite{31,32}. Utilizing the similar phase transition characteristics of RN-AdS BH and vdW systems, Wei $et~al.$ rewrote the Ruppeiner curvature describing the microscopic property of BH systems to the normalized Ruppeiner scalar curvature. They used the new normalized scalar curvature to analyze the microstructure of the RN-AdS BH and vdW system. Note that this promotion eliminates the difficulty of analogy with classical systems caused by inconsistent heat capacity behavior. The microcosmic consistency of the RN-AdS BH and vdW system shows that the interaction of attraction is dominant in the microcosmic, and the normalized curvature at the critical point has recognizable geometric behavior. In particular, for some small RN-AdS BH systems, the repulsive interaction is dominant in the micro, which is different from the vdW system \cite{33,34}.

\par
Thermodynamic properties of BHs in the extended phase space can be used as a window to explore BHs and their properties under different gravitational frameworks. For the BH in massive gravity theory, it is worth studying how gravitons with mass affect the thermodynamic properties of the whole system and how this effect is shielded. Cai $et~al.$ presented a class of charged AdS BH solutions in an $(n+2)$-dimensional massive gravity and investigated the thermodynamics and phase structure from the view of grand canonical and canonical ensembles. They found that the behaviors of phase transition caused by charge are slightly regulated by the graviton. Interestingly, in the higher-dimensional case [$ n+2\geq 5$], the mass of graviton can cause the phase transition of large-small BH without charge \cite{35}. When the space-time dimension is larger than four, the authors found that proper adjustment of the massive potential can provide a repulsive effect similar to the electric potential, so only the graviton with mass can form a second-order phase transition in dimension \cite{36}.

\par
Based on these researches, it is unclear whether the graviton with mass that behaves as scalar interaction will make the BH behave similarly charged in the four-dimensional case. This paper focuses on this issue. One of the aims of the present work is to extend the above formalism of the extended phase space to the case of BHs arising in massive gravity. The effect of different $\lambda$ parameters of massive gravity theory on phase transition caused by scalar charge is investigated. We compare these BHs with RN-AdS BH in terms of phase transition grade, phase transition critical exponents, and micro behaviors. Our results imply the phase transition of BHs in massive gravity is almost consistent with the phase transition of RN-AdS BH, the consistency of this may be that the interaction between BH and graviton with mass is also accompanied by the release of the degenerate state, just like the charge. The paper is organized as follows: In section \ref{sec:2}, the thermodynamics of BHs in massive gravity was presented. In section \ref{sec:3}, the phase transition grade was checked and the critical exponents were calculated with the macroscopic method and microscopic approach. In section \ref{sec:4}, Ruppeiner geometry was constructed to reveal the microstructure of the thermodynamic space of BHs. In section \ref{sec:5}, the conclusion was given as an end.

\section{\textbf{Thermodynamics of BH in massive gravity}}
\label{sec:2}
In massive gravity theory, the four-dimensional action is \cite{11}
\begin{equation}
S=\int {\rm d}^4x\sqrt{-g}(-M^2_{pl}R+\mathcal{L}_{m}+\Lambda^4 F),
\label{2-1}
\end{equation}
where $g$ is determinant of the metric tensor, $R$ is the Ricci scalar, $\mathcal{L}$ is Lagrange density of the minimally coupled ordinary matter, the third term describes four scalar fields $\phi^0$, $\phi^i$ whose space-time dependent vacuum expectation values $\Lambda$ break spontaneously the Lorentz symmetry. $F$ is an arbitrary function of metric components, their derivatives and coordinates $t$, $x^{i}(i=1,2,3)$ itself, so it can be set as the function of two particular combinations of the derivatives of Goldstone fields. The static and spherically symmetric metric is
\begin{equation}
{\rm d}s^2=-h(r){\rm d}t^2+{\rm d}r^2/h(r)+r^2({\rm d}\theta^2+sin^2\theta {\rm d}\phi^2),
\label{2-2}
\end{equation}
where $h(r)$ is the metric potential, its \cite{11}
\begin{equation}
h(r)=1-\frac{2M}{r}-\gamma \frac{Q^2}{r^{\lambda}}-\frac{\Lambda r^2}{3},
\label{2-3}
\end{equation}
in which $M$ is the BH mass, $Q$ is the scalar charge of BH. The parameter $\lambda$ is state constant ($\lambda \geq 2$), which comes from massive gravity theory. $\Lambda$ is vacuum expectation value, which carries attribute of cosmological constant. $\gamma$ is a free parameter, satisfying $\gamma= \pm 1$. We only focus on the case of $\gamma= -1$ in this analysis because the geometry of BHs in this situation is similar to the RN-AdS BH. Note that $\Lambda$ is considered as the thermodynamic pressure in the extended phase space, i.e. $P=-{\Lambda}/{8\pi}$ \cite{37}. The radius of the event horizon $r_{\rm +}$ is the largest root of $h(r_{\rm +})=0$. According to Eq.(\ref{2-3}), one can obtain that the BH mass,
\begin{equation}
M=\frac{3r_{+}^{\lambda} + 8P\pi r_{+}^{2+\lambda}-3Q^2\gamma}{6r_{+}^{\lambda-1}}.
\label{2-4}
\end{equation}
Thinking of the area law, the BH entropy is $S=\pi r_{+}^2$. Subsequently, the BH mass $M$ can be rewritten as a function of the entropy $S$, pressure $P$ and scalar charge $Q$, i.e.
\begin{equation}
M=\frac{3\sqrt{S}+8PS^{3/2}-3\pi^{\lambda/2}Q^2S^{1/2-\lambda/2}\gamma}{6\sqrt{\pi}}.
\label{2-5}
\end{equation}
Using the first law of BH thermodynamics, all the thermodynamic variables of the BH under massive gravity framework are obtained, we have
\begin{equation}
T=\Bigg(\frac{\partial M}{\partial S}\Bigg)_{\rm Q,P},~~~~\Phi=\Bigg(\frac{\partial M}{\partial Q}\Bigg)_{\rm S,P},~~~~V=\Bigg(\frac{\partial M}{\partial P}\Bigg)_{\rm S,Q}.
\label{2-6}
\end{equation}
Thus, the BH temperature reads
\begin{equation}
T=\frac{S^{\frac{1+\lambda}{2}}+8PS^{\frac{3+\lambda}{2}}+\pi^{\frac{\lambda}{2}}Q^2\sqrt{S}\gamma(\lambda-1)}{4\sqrt{\pi}S^{\frac{2+\lambda}{2}}},
\label{2-7}
\end{equation}
the scalar charge conjugate quantity is
\begin{equation}
\Phi=-\pi^{\frac{\lambda-1}{2}}QS^{\frac{1-\lambda}{2}}\gamma,
\label{2-8}
\end{equation}
and the thermodynamic volume is
\begin{equation}
V=\frac{4S^{\frac{3}{2}}}{3\sqrt{\pi}}.
\label{2-9}
\end{equation}
The isobaric heat capacity and Gibbs free energy are
\begin{equation}
C_{\rm P,q}= T\Bigg(\frac{\partial S}{\partial T}\Bigg)_{\rm P,q}=\frac{2\pi r^2\Big(r^{\lambda}(1+8\pi Pr^2)+Q^2\gamma(\lambda-1)\Big)}{r^{\lambda}(8\pi Pr^2-1)-Q^2\gamma(\lambda^2-1)},
\label{2-10}
\end{equation}
and
\begin{equation}
G=M-TS=\frac{r}{3}-\frac{\pi r^2T}{3} -\frac{Q^2\gamma(3r^{2\lambda}+\lambda-1)}{6r^{\lambda-1}}.
\label{2-11}
\end{equation}
According to the formulas of heat capacity and Gibbs function, the heat capacity $C_{\rm P}$ as a function of horizon radius $r_{+}$, and Gibbs free energy as a function of temperature $T$ under several representative values of state constant $\lambda$ are shown in Fig.1. We can see that $C_{\rm P}$ has a mutation at a specific radius, indicating that the phase transition of BH under massive gravity framework is a high-order phase transition. It is also found that $G$ functions have swallowtail behavior.

\begin{center}
\includegraphics[width=7cm,height=5.5cm]{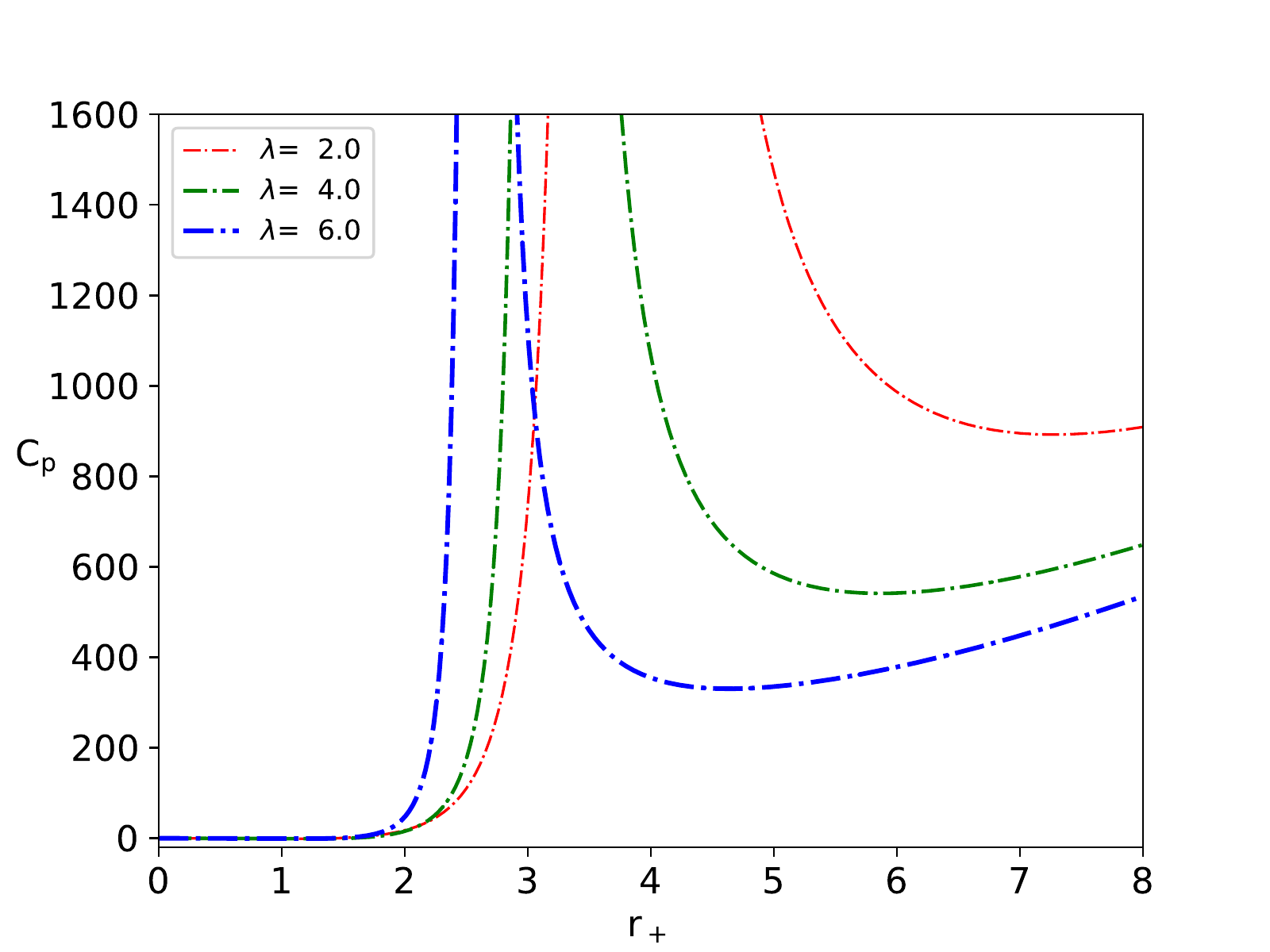}
\hspace{0.8cm}
\includegraphics[width=7cm,height=5.5cm]{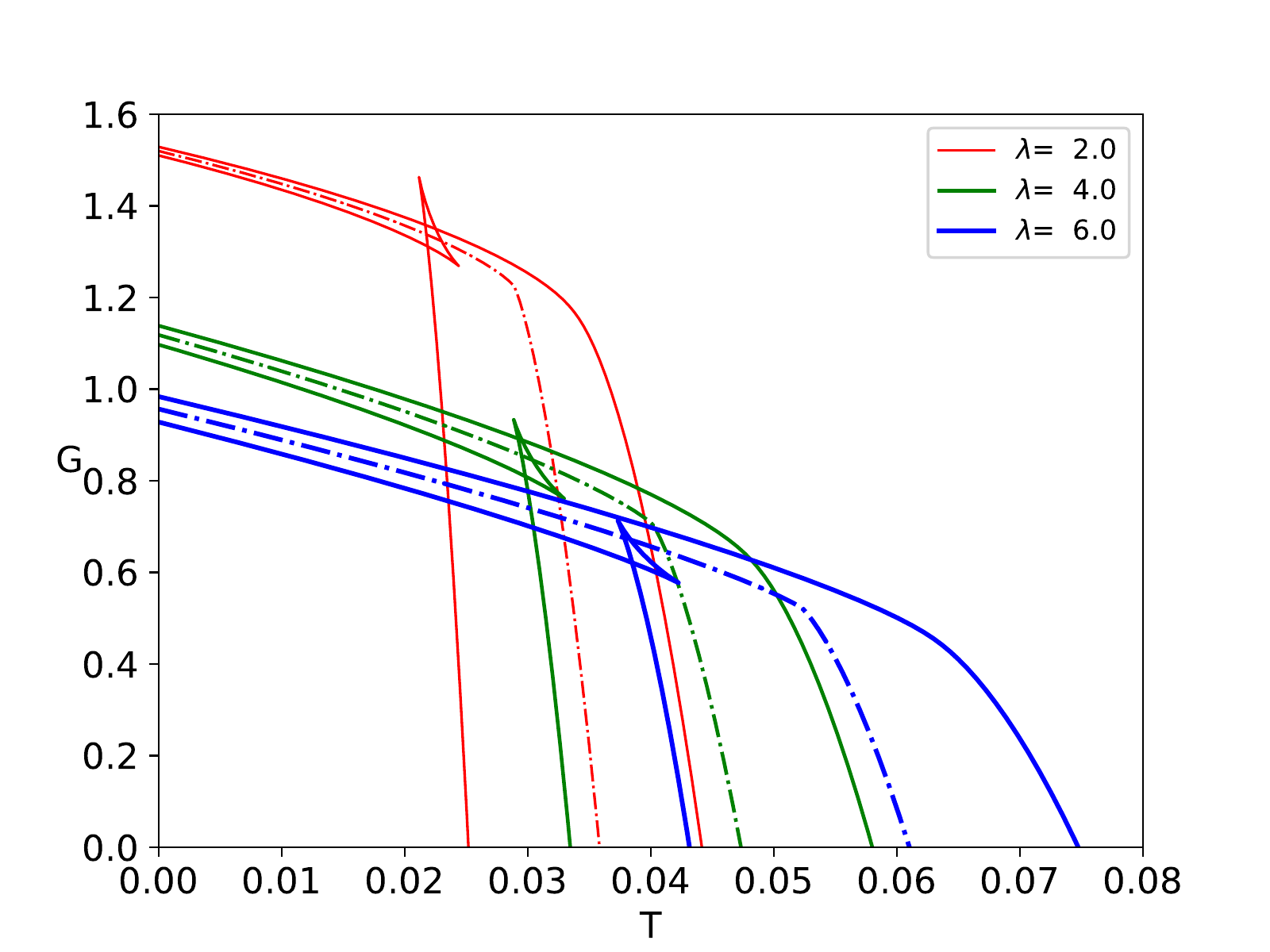}
\parbox[c]{15.0cm}{\footnotesize{\bf Fig~1.}  
Left: heat capacity $C_{P}$ as a function of the horizon radius $r_{+}$. Right: Gibbs free energy $G$ as a function of the BH temperature $T$. The red, green and blue represent $\lambda$ parameters as $2,~4,~6$, respectively.}
\label{fig1}
\end{center}

\par
To make the following calculation more convenient and also can reflect the influence of parameter $\lambda$, we simplify the forms of parameters of BH in massive gravity as follows,
\begin{eqnarray}
\widetilde{m}=aM,~~\widetilde{r}_{+}=ar_{+},~~\widetilde{p}=\frac{P}{a^2},~~\widetilde{q}=\frac{Q}{a},~~\widetilde{t}=\frac{T}{a},
\label{2-12}
\end{eqnarray}
where $a$ satisfies ${-a^4 \gamma}/{r^{\lambda-2}}=1$. Thus, the equation of state of the BH under massive gravity framework can be written as
\begin{equation}
\widetilde{p}=\frac{\widetilde{t}}{2 \widetilde{r}_{+}}-\frac{1}{8\pi \widetilde{r}_{+}^2}+\frac{\widetilde{q}^2(\lambda-1)}{8\pi \widetilde{r}_{+}^4}.
\label{2-13}
\end{equation}
Utilizing BH critical conditions $(\frac{{\partial P}}{{\partial r_{+}}})=(\frac{{\partial^2 P}}{{\partial r_{+}^2}})=0$ \cite{22}, the simplify critical thermodynamic quantities are obtained, we have
\begin{eqnarray}
\widetilde{t}_{\rm c}&=&\frac{\lambda}{2\pi(\lambda+1)\widetilde{r}_{\rm c}},\\
\label{2-14}
\widetilde{p}_{\rm c}&=&\frac{\lambda}{8\pi(\lambda+2)\widetilde{r}_{\rm c}^2},\\
\label{2-15}
\widetilde{r}_{\rm c}^2&=&\frac{\widetilde{q}_{\rm c}^2(\lambda+2)(\lambda+1)(\lambda-1)}{2}.
\label{2-16}
\end{eqnarray}
The critical universal constant is defined as
\begin{equation}
\varepsilon \equiv \frac{P_{\rm c} \upsilon_{\rm c}} {T_{\rm c}},
\label{2-17}
\end{equation}
where the $\upsilon$ is the specific volume, satisfying $\upsilon=2r_{+}$. Based on Eq.(\ref{2-12}), one can obtain $\varepsilon= \widetilde{\varepsilon}= \widetilde{p}_{\rm c}\widetilde{\upsilon}_{\rm c}/\widetilde{t}_{\rm c}$, where $\widetilde{\upsilon}$ is the simplified specific volume, hence, the critical universal constant is
\begin{equation}
\varepsilon = \frac{4(\lambda+1)}{8(\lambda+2)}.
\label{2-18}
\end{equation}
It deviates slightly from the vdW system, but it still belongs to this category. Setting parameter $\lambda=2$, one can get
\begin{equation}
\varepsilon_{\rm \lambda=2}=\frac{3}{8}.
\label{2-19}
\end{equation}
The result is same as vdW system and RN-AdS BH, and the larger parameter $\lambda$ leads to the stronger critical universal constant, and it has obvious limit $3/8< \varepsilon <1/2$.

\par
By introducing simplified thermodynamic quantities, we review the thermodynamic properties of the BH under the massive gravity framework in this section. It is found that the isobaric heat capacity and Gibbs free energy show the unique behaviors of a second-order phase transition. However, the more rational judgment of the phase transition grade in classical thermodynamics depends on the Clausius-Clapeyron-Ehrenfest equations. Hence, the phase transition grade of the BH is investigated from macroscopic and microcosmic perspectives in next section.

\section{\textbf{Phase transition grade of the BHs under massive gravity framework}}
\label{sec:3}
\subsection{Macroscopic point of view: Ehrenfest equations}
\label{sec:3-1}
\par
According to the Gibbs function continuity in the classical thermodynamics process, the phase transition grade of the system can be subdivided into the first-order and high-order phase transition. The Clausius-Clapeyron-Ehrenfest equations provide a strong guarantee for clearly judging the phase transition grade, the first-order phase transition follows Clausius-Clapeyron equations, and the second-order phase transition satisfies Ehrenfest equations.

\par
The Ehrenfest equations can be written as
\begin{eqnarray}
\label{3-1-1}
-\Bigg(\frac{\partial P}{\partial T}\Bigg)_{\rm S,Q} &=& \frac{1}{V T}\frac{C_{\rm P2}-C_{\rm P1}}{\alpha_{\rm 2}-\alpha_{\rm 1}} = \frac{\Delta C_{\rm P}}{V T \Delta \alpha},\\
\label{3-1-2}
-\Bigg(\frac{\partial P}{\partial T}\Bigg)_{\rm V,Q} &=& \frac{\alpha_{\rm 2}-\alpha_{\rm 1}}{\kappa_{\rm T2}-\kappa_{\rm T1}} = \frac{\Delta \alpha}{\Delta \kappa_{\rm T}},
\end{eqnarray}
where the subscript $1$ and $2$ represent the phase $1$ and the phase $2$ near the critical point, $\alpha$ and $\kappa$ correspond to the expansion coefficient and compressibility coefficient,
\begin{equation}
\label{3-1-3}
\alpha=-\frac{1}{V}\Bigg(\frac{\partial V}{\partial T}\Bigg)_{\rm P,Q},~~~~~\kappa_{\rm T}=-\frac{1}{V}\Bigg(\frac{\partial V}{\partial P}\Bigg)_{\rm T,Q}.
\end{equation}
Utilizing Eq.(\ref{2-12}), the simplified forms of the heat capacity, the expansion coefficient and the compressibility coefficient are obtained, we have
\begin{equation}
\label{3-1-4}
\widetilde{C}_{\rm P}=a^2C_{\rm P},~~~~\widetilde{\alpha} = a \alpha,~~~~\widetilde{\kappa}_{\rm T} = a^2\kappa_{\rm T}.
\end{equation}
Hence, the simplified forms of Ehrenfest equations can be rewritten as
\begin{equation}
\label{3-1-5}
-\Bigg(\frac{\partial \widetilde{P}}{\partial \widetilde{T}}\Bigg)_{\rm \widetilde{S},\widetilde{Q}} = \frac{\Delta \widetilde{C}_{\rm \widetilde{P}}}{\widetilde{V} \widetilde{T} \Delta \widetilde{\alpha}},~~~-\Bigg(\frac{\partial \widetilde{P}}{\partial \widetilde{T}}\Bigg)_{\rm \widetilde{V},\widetilde{Q}} = \frac{\Delta \widetilde{\alpha}}{ \Delta \widetilde{\kappa}_{\rm T}}.
\end{equation}

\par
For the case of the BH under massive gravity framework, the left side of the Ehrenfest equations read as
\begin{equation}
\label{3-1-6}
-\Bigg(\frac{\partial \widetilde{P}}{\partial \widetilde{T}}\Bigg)_{\rm \widetilde{S},\widetilde{Q}} = -\Bigg(\frac{\partial \widetilde{P}}{\partial \widetilde{T}}\Bigg)_{\rm \widetilde{V},\widetilde{Q}} = \frac{1}{2\widetilde{r}}.
\end{equation}
The right side of the Ehrenfest equations as
\begin{eqnarray}
\label{3-1-7}
&\frac{\Delta \widetilde{C}_{\rm \widetilde{P}}}{\widetilde{V} \widetilde{T} \Delta \widetilde{\alpha}} = \frac{1}{a V T}\frac{C_{\rm P2}-C_{\rm P1}}{\alpha_{2}-\alpha_{1}}\nonumber\\
&=\frac{8\pi\widetilde{p}_{\rm 2}\widetilde{r}_{\rm 2}\widetilde{x}^{\lambda+2}(\widetilde{x}-1)-\widetilde{x}^{\lambda}(\widetilde{x}^3-1)
+\widetilde{q}_{\rm 2}(\widetilde{x}^{\lambda+3}-1)(\lambda^2-1)}{8\pi\widetilde{p}_{\rm 2}\widetilde{r}_{\rm 2}\widetilde{x}^{\lambda+2}(\widetilde{x}^2-1)
-\widetilde{x}^{\lambda}(\widetilde{x}^4-1)+\widetilde{q}_{\rm 2}(\widetilde{x}^{\lambda+4}-1)(\lambda^2-1)}\frac{1}{2\widetilde{r}_{\rm 2}},
\end{eqnarray}
and
\begin{eqnarray}
\label{3-1-8}
&\frac{\Delta \widetilde{\alpha}}{ \Delta \widetilde{\kappa}_{\rm T}} = \frac{\alpha_{\rm 2}-\alpha_{\rm 1}}{a(\kappa_{\rm T2}-\kappa_{\rm T1})}\nonumber\\
&=\frac{8\pi\widetilde{p}_{2}\widetilde{r}_{2}\widetilde{x}^{\lambda+2}(\widetilde{x}^2-1)
-\widetilde{x}^{\lambda}(\widetilde{x}^4-1)
+\widetilde{q}_{2}(\widetilde{x}^{\lambda+4}-1)(\lambda^2-1)}{8\pi\widetilde{p}_{2}\widetilde{r}_{2}\widetilde{x}^{\lambda+2}(\widetilde{x}^3-1)
-\widetilde{x}^{\lambda}(\widetilde{x}^5-1)+\widetilde{q}_{2}(\widetilde{x}^{\lambda+5}-1)(\lambda^2-1)}\frac{1}{2\widetilde{r}_{2}},
\end{eqnarray}
where the $\widetilde{x} \equiv \widetilde{r}_{1}/\widetilde{r}_{2}$, and the $\widetilde{x} \rightarrow 1$ at the critical point. Thus, one can get
\begin{eqnarray}
\label{3-1-9}
\lim_{\widetilde{x}\to1}\Bigg(\frac{1}{a V T}\frac{C_{\rm P2}-C_{\rm P1}}{\alpha_{2}-\alpha_{1}}\Bigg) = \lim_{\widetilde{x}\to1} \Bigg(\frac{\alpha_{2}-\alpha_{1}}{a(\kappa_{\rm T2}-\kappa_{\rm T1})}\Bigg) = \frac{1}{2\widetilde{r}}.
\end{eqnarray}
As a result, we can obtain that the BHs under the massive gravity framework satisfies Ehrenfest equations near the critical point, implying that the phase transition of the BH is {\color{blue}a} second-order phase transition. It is also found that the parameter $\lambda$ does not affect the BH phase transition grade. Furthermore, the second-order phase transition can be subdivided into the standard phase transition and glassy phase transition by the Prigogine-Defay (PD) ratio. The case of $PD = 2 \sim 5$ is for the glassy phase transition. In our analysis, the PD ratio of BHs in massive gravity is
\begin{equation}
\label{3-1-10}
\Pi=\frac{(C_{\rm P2}-C_{\rm P1})(\kappa_{\rm T2}-\kappa_{\rm T1})}{TV(\alpha_{2}-\alpha_{1})^2}=1.
\end{equation}
It is means that the phase transition of these BHs is a standard second-order phase transition.

\par
We also derive the phase transition critical exponents for the BH. The critical exponents can be given by
\begin{eqnarray}
C_{V}&=&T\Big(\frac{\partial S}{\partial T}\Big)_{V}\propto|\iota|^{-\alpha},~~~~\eta=V_{2}-V_{1}\propto(-\iota)^{\beta},\nonumber\\
\label{3-1-11}
\kappa_{T}&=&-\frac{1}{V}\Big(\frac{\partial V}{\partial P}\Big)\propto|\iota|^{-\gamma},~~~P-P_{c}\propto|V-V_{c}|^{\delta}.
\end{eqnarray}
The first critical exponent $\alpha$ can be identified as 0 ($\alpha = 0$) since the isovolumetric heat capacity $C_{\rm V}=0$. The other critical exponents are derived from the reduced thermodynamic quantities
\begin{equation}
\label{3-1-12}
p=\frac{P}{P_{\rm c}},~~~~\nu=\frac{\upsilon}{\upsilon_{\rm c}},~~~~\tau=\frac{T}{T_{\rm c}},
\end{equation}
where $p$, $\nu$, and $\tau$ are the reduced pressure, reduced volume, and reduced temperature, respectively. Considering the simplified forms, the above equations can be written as
\begin{eqnarray}
\label{3-1-13}
\frac{\widetilde{p}}{\widetilde{p}_{\rm c}}=\frac{a_{\rm c}^2}{a^2}p,~~
\quad\frac{\widetilde{\upsilon}}{\widetilde{\upsilon}_{\rm c}}=\frac{a}{a_{\rm c}}\nu,~~~~~
\frac{\widetilde{t}}{\widetilde{t}_{\rm c}}=\frac{a_{\rm c}}{a}\tau.
\end{eqnarray}
The reduced state equation is
\begin{eqnarray}
\label{3-1-14}
p=\frac{2(\lambda+2) \tau }{(\lambda+1)\nu }-\frac{\lambda+2}{\lambda {\nu}^{2}}+\frac{2}{\lambda(\lambda+1){\nu}^{\lambda+2}}.
\end{eqnarray}
By introducing the quantities
\begin{equation}
\label{3-1-15}
{\tau}={\iota}+1,\qquad {\nu}=({\omega}+1)^{1/3},
\end{equation}
one can get
\begin{equation}
\label{3-1-16}
{p}=1+A_{\lambda}{\iota}-B_{\lambda}{\iota}{\omega}-C_{\lambda}{\omega}^{3}+O({\iota}{\omega}^{2},{\omega}^{4}),
\end{equation}
where
\begin{eqnarray}
\label{3-1-17}
A_{\lambda}=\frac{2\lambda+4}{\lambda+1},~~B_{\lambda}=\frac{4\lambda+8}{\lambda^{2}+\lambda},~~
C_{\lambda}=\frac{\lambda+2}{3(\lambda+1)^3}.\nonumber
\end{eqnarray}
The differential of reduced pressure is
\begin{equation}
\label{3-1-18}
dp=-(B_{\lambda}{\iota}+3C_{\lambda}{\omega}^{2})d{\omega}.
\end{equation}
With Maxwell's equal area law, the `volume' of small BH $\omega_{1}$ and large BH $\omega_{2}$ have the following relationship
\begin{equation}
\label{3-1-19}
\int_{\omega_{1}}^{\omega_{2}}\omega dp=0,
\end{equation}
this equation has unique nontrivial solution, that is
\begin{equation}
\label{3-1-20}
\omega_{1}=-\omega_{2}=\sqrt{\frac{B_{\lambda}}{C_{\lambda}}(-\iota)}.
\end{equation}
Considering above formula, other critical exponents are derived. For the case of the second critical exponent, we have
\begin{equation}
\label{3-1-21}
\eta=V_{2}-V_{1}=V_{c}(\omega_{2}-\omega_{1})=2V_{\rm c}\omega_{2},
\end{equation}
where $\omega_{2} \propto\sqrt{-\iota}$. Therefore, the second critical exponent $\beta = 1/2$.

\par
The isothermal compressibility coefficient is
\begin{equation}
\label{3-1-22}
\kappa_{\rm T}=-\frac{1}{V}\Bigg(\frac{\partial V}{\partial P}\Bigg)=-\frac{1}{P_{\rm c}(\omega+1)}\frac{{\rm d}\omega}{{\rm d}p}\propto\frac{1}{B_{\lambda}\iota},
\end{equation}
hence the third critical exponent $\gamma=1$. Finally, the following formula is naturally derived when $\iota=0$,
\begin{equation}
\label{3-1-23}
p-1=-C_{\lambda}\omega^{3},
\end{equation}
implying that the fourth critical exponent $\delta=3$.

\par
From the perspective of macro thermodynamics, we prove the phase transition of the BHs under the massive gravity framework satisfies Ehrenfest equations in this subsection, implying that the phase transition grade of this BH is a standard second-order phase transition. Meanwhile, we obtain that the critical exponents of this BH are $(\alpha,\beta,\gamma,\delta)=(0,~1/2,~1,~3)$, which is similar to the vdW system. We also found that the parameter $\lambda$ does not affect the BH phase transition grade.

\subsection{Microcosmic point of view: Landau continuous phase transition}
\label{sec:3-2}
\par
Taking the ferromagnetic molecules as an example, Landau's continuous phase transition theory reports the phase transition process by revealing the changes in order degree and spatial symmetry of the system. Concretely, a disordered and high symmetry system appears above the critical point, but below the critical point is an orderly degree and low symmetry situation. These two scenarios indicate that the critical point is a link used to combine the system with phase transition occurs and the system without phase transition occupation. According to Landau's theory, an order parameter can be constructed to describe the above characteristics, showing a transition from zero to non-zero when the temperature gradually decreases from the critical point to below the critical point.

\par
Based on Ref.\cite{38}, a BH may be full of BH molecules like air molecules for carrying the BH entropy microscopic degree freedoms, and its microstructure is similar to the ordinary thermodynamic system. As a result, we can analyze the phase transition behaviors of the BH under the massive gravity framework from the microcosmic point of view. When the BH temperature is higher than the critical temperature, BH molecules collide violently since they have high kinetic energy, which shows the high symmetry and disordered of this case. While the BH temperature is lower than the critical temperature, the BH molecules lose activity at low temperatures, representing the low symmetry and orderly. There are two phases near the critical point, the high potential phase has lower symmetry and the low phase has higher symmetry. In these three cases, the order parameter is directly proportional to the order degree of the BH molecules and inversely proportional to the symmetry of the BH system. On the other hand, the scalar charge plays a crucial role in the phase transition, which should be considered in the BH microstructure theory. The order parameter of the BH system is the production of degeneracy release of the BH molecules caused by the scalar charge. Therefore, we choose the potential $\Psi$ as the order parameter to analyze the phase transition of the BH in massive gravity from the microscopic view.

\par
If the temperature is fixed and lower than the critical temperature, the simplified potentials of any two phases are
\begin{eqnarray}
\label{3-2-1}
\widetilde{\phi}_{1}=\frac{\widetilde{q}_{1}}{\widetilde{r}_{1}},~~~~~~\widetilde{\phi}_{2}=\frac{\widetilde{q}_{2}}{\widetilde{r}_{2}},
\end{eqnarray}
where $\widetilde{r}_{1}$ stands for the radius of the strong potential phase $\widetilde{\phi}_{1}$ with low symmetry, high-order degree, and non-zero order-parameter, $\widetilde{r}_{2}$ represents the radius of the weak potential phase $\widetilde{\phi}_{2}$ with relatively lower symmetry, higher-order degree, and non-zero order-parameter. The order-parameter can be expressed by the normalized potential difference between the high potential phase and low potential phase, we have
\begin{equation}
\label{3-2-2}
\widetilde{\Psi}(\widetilde{t})=\frac{\widetilde{\phi}_{1}-\widetilde{\phi}_{2}}{\widetilde{\phi}_{c}}=\frac{\widetilde{\phi}_{2}}{\widetilde{\phi}_{c}}\Bigg(\Big(\frac{1}{\widetilde{x}}\Big)^{\frac{\lambda+2}{4}}-1\Bigg).
\end{equation}

\par
It is shown that an order parameter is a function of the BH temperature, so the Gibbs free energy can be rewritten as a function of temperature and order parameter. Note that when we examine the phase transition near the critical point, there will be the relationship $\Psi(T)=\widetilde{\Psi}(\widetilde{t})$. According to Landau's theory, the order parameter as a small quantity near the critical point leads to the Gibbs function can be expanded into a power series of the order parameter. It is worth noting that the symmetry of the phase transition ($\phi \rightleftharpoons -\phi$) is also passed to the Gibbs function, i.e. the odd power terms of $\Psi$ vanish. Hence, the Gibbs function can be written as
\begin{equation}
\label{3-2-3}
G(T,\Psi)=G_{0}(T)+\frac{1}{2}A(T)\Psi^{2}+\frac{1}{4}B(T)\Psi^{4}+O(T,\Psi^{6})+...,
\end{equation}
where the $G_{0}(T)$ is the free energy of $\Psi(T)=0$, A and B are parameters depending on the system temperature. Considering the minimum condition of stable equilibrium Gibbs function, one can get
\begin{eqnarray}
\label{3-2-4}
\frac{\partial G}{\partial\Psi}=\Psi(A+B\Psi^{2})=0,\qquad\frac{\partial^2 G}{\partial\Psi^2}=A+3B\Psi^2>0,
\end{eqnarray}
the above requirements give the three solutions as
\begin{eqnarray}
\label{3-2-5}
\Psi=0,\qquad\Psi=\pm\sqrt{-\frac{A}{B}}.
\end{eqnarray}
The solution $\Psi=0$ meets $T = T_{c}$, which represents BH molecules in the disordered state. With the decrease in temperature, the state of the BH system will change from disorder to order. Once out of the state represented by the critical point, the order parameter $\Psi= \pm\sqrt{-A/B}$, it realizes the transition from zero to non-zero. Assuming that $A$ can be written as a function of $T$, and $B$ is a constant, we have
\begin{eqnarray}
\label{3-2-6}
A(T)=A_{0}\Bigg(\frac{T-T_{c}}{T_{c}}\Bigg)=A_{0}\iota \qquad A_{0}>0,
\end{eqnarray}
if the $T<T_{c}$, $A(T)$ is a negative number, therefore, $B$ is a positive number. According to Eqs. (\ref{3-2-5}) and (\ref{3-2-6}), the $\Psi$ can be written as
\begin{equation}
\label{3-2-7}
\Psi=\left\{
\begin{array}{cl}
0 & for\quad \iota>0,\\
\\
\pm\sqrt\frac{A_{0}}{B}(-\iota)^{\frac{1}{2}} & for\quad \iota<0.
\end{array} \right.
\end{equation}
Based on above equation, one can get the second critical exponent $\beta = 1/2$. The Gibbs function can be rewritten as
\begin{equation}
\label{3-2-8}
G(T,\Psi)=\left\{
\begin{array}{cl}
G_{0}(T) & for\quad T>T_{c},\\
\\
G_{0}(T)-\frac{A_{0}^2}{4B}(\frac{T-T_{c}}{T_{c}}) & for\quad T<T_{c}.
\end{array} \right.
\end{equation}
Considering the heat capacity $C=-T(\partial^2G/\partial T^2)$, the difference in heat capacity between phase $1$ and phase $2$ is
\begin{equation}
\label{3-2-9}
C(\iota<0)\mid_{\iota=0}-C(\iota>0)\mid_{\iota=0}=\frac{A_{0}^2}{2BT_{c}},
\end{equation}
one can observe that the heat capacity of the ordered phase ($\iota<0$) is bigger than that of the disordered phase ($\iota>0$), and the sudden change of heat capacity at $\iota=0$ is limited, it follows that the first critical exponent $\alpha$ is 0.

\par
Considering a constant pressure, the total differential of the BH Gibbs function reads as
\begin{equation}
\label{3-2-10}
{\rm d}G=-S{\rm d}T-Q{\rm d}\Psi.
\end{equation}
Utilizing Eqs.(\ref{3-2-3}) and (\ref{3-2-10}), we have
\begin{eqnarray}
\label{3-2-11}
-Q=\Big(\frac{\partial G}{\partial \Psi}\Big)=\Psi(A+B\Psi^{2}),\\
-\Big(\frac{\partial \Psi}{\partial Q}\Big)=\frac{1}{A+3B\Psi^{2}}\left\{
\begin{array}{cl}
\frac{1}{a_{0}}\iota^{-1} & for\quad \iota>0,\\
\\
\frac{1}{2a_{0}}(-\iota)^{-1}& for\quad \iota<0.
\end{array} \right.
\end{eqnarray}
Therefore, the third critical exponent $\gamma$ is 1. The relation $Q\propto\Psi^3$ shows that the last critical exponent $\delta$ is 3.

\par
According to the Landau continuous phase transition, this subsection analyzed the phase transition of the BH in massive gravity from the microcosmic point of view. Based on the BH molecules model, the critical exponents consistent with the macro analysis are obtained, i.e. $(\alpha,~\beta,~\gamma,~\delta)=(0,~1/2,~1,~3)$. We also found that the $\lambda$ parameter does not significantly affect the phase transition characteristics. These results suggest that the phase transition of the BH in massive gravity is a second-order phase transition from various perspectives, and it is still a vdW-like phase transition. To further understand the phase structure of the BH under the massive gravity framework, especially, the influence of parameter $\lambda$ of massive gravity theory on the phase near the critical point, we use Ruppeiner geometry to reveal the feature of the BH phases transition in next section.

\section{\textbf{Ruppeiner geometry for the BHs under massive gravity framework}}
\label{sec:4}
The Ruppeiner geometry may serve as a probe for the microstructure structure of a thermodynamic system. The geometric manifold of a thermodynamic space is composed of microstates, further{,} the microcosmic structure depends on the entropy. For the different microcosmic states, the entropy difference is inversely proportional to the fluctuation probability. Therefore, the thermodynamic characteristics of the system can be presented by the Ruppeiner geometry, which reveals the microscopic behaviors of the system. In Refs. \cite{33,34}, the normalized scalar curvature ${\rm R}_{\rm N}$ is defined as ${\rm R}_{\rm N} \equiv C_{\rm V} R$, which is used to replace scalar curvature $R$ to analyze the microstructure of RN-AdS BH and vdW system. Based on the isovolumetric heat capacity of the vdW system $C_{\rm V}= \frac{3}{2}k_{\rm B}$ is of order $10^{-23}$, it is considered that $C_{\rm V}\rightarrow0$ to avoid the behavior of $C_{\rm V}=0$ in the expanded phase space. We can find that if the $C_{\rm V}$ is a constant, ${\rm R}_{\rm N}$ can well inherit the divergence of $R$, and if $C_{\rm V}$ approaches zero, the divergence of ${\rm R}_{\rm N}$ means that there is an $R$ with higher-order divergence. Therefore, it is accurate to use ${\rm R}_{\rm N}$ for microscopic feature analysis when comparing BHs in the expanded phase space with classical systems. The normalized scalar curvature ${\rm R}_{\rm N}$ is given by \cite{33,34}
\begin{equation}
\label{4-1}
{\rm R}_{\rm N}=\frac{(\partial_{\rm V}P)^2-T^2(\partial_{\rm V,T})^2+2T^2(\partial_{\rm V}P)(\partial_{\rm V,T,T})}{2(\partial_{\rm V}P)^2}.
\end{equation}
By introducing Eq.(\ref{3-1-14}), one can get
\begin{equation}
\label{4-2}
{\rm R}_{\rm N}=-\frac{(\nu^{\lambda}(\lambda+1)-1)(\nu^{\lambda}(\lambda(2\tau\nu-1)-1)+1)}{2(\nu^{\lambda}(\lambda(1-\tau\nu)+1)-1)^2}.
\end{equation}
Based on this equation, the normalized scalar curvature ${\rm R}_{\rm N}$ as a function of the $\nu$ with a fixed temperature is shown in Fig.2. We can see that the two divergence points of ${\rm R}_{\rm N}$ tend to coincide when the BH temperature is equal to the $T_{\rm c}$, and the sign of ${\rm R}_{\rm N}$ has changed from positive to negative. With the decrease of temperature, the distance between the two divergence points gradually becomes farther, and the transformation of the sign of ${\rm R}_{\rm N}$ is still once. When the temperature drops to a certain range, the distance between divergence points is getting larger and larger, but the sign of ${\rm R}_{\rm N}$ changed three times (first column of Fig.2). It is found that $T>T_{\rm c}$, $T=T_{\rm c}$, and $T<T_{\rm c}$ correspond to the cases of zero divergence point, one divergence point, and two divergence points of the ${\rm R}_{\rm N}$, respectively.

\begin{center}
\includegraphics[width=9.5cm,height=6.5cm]{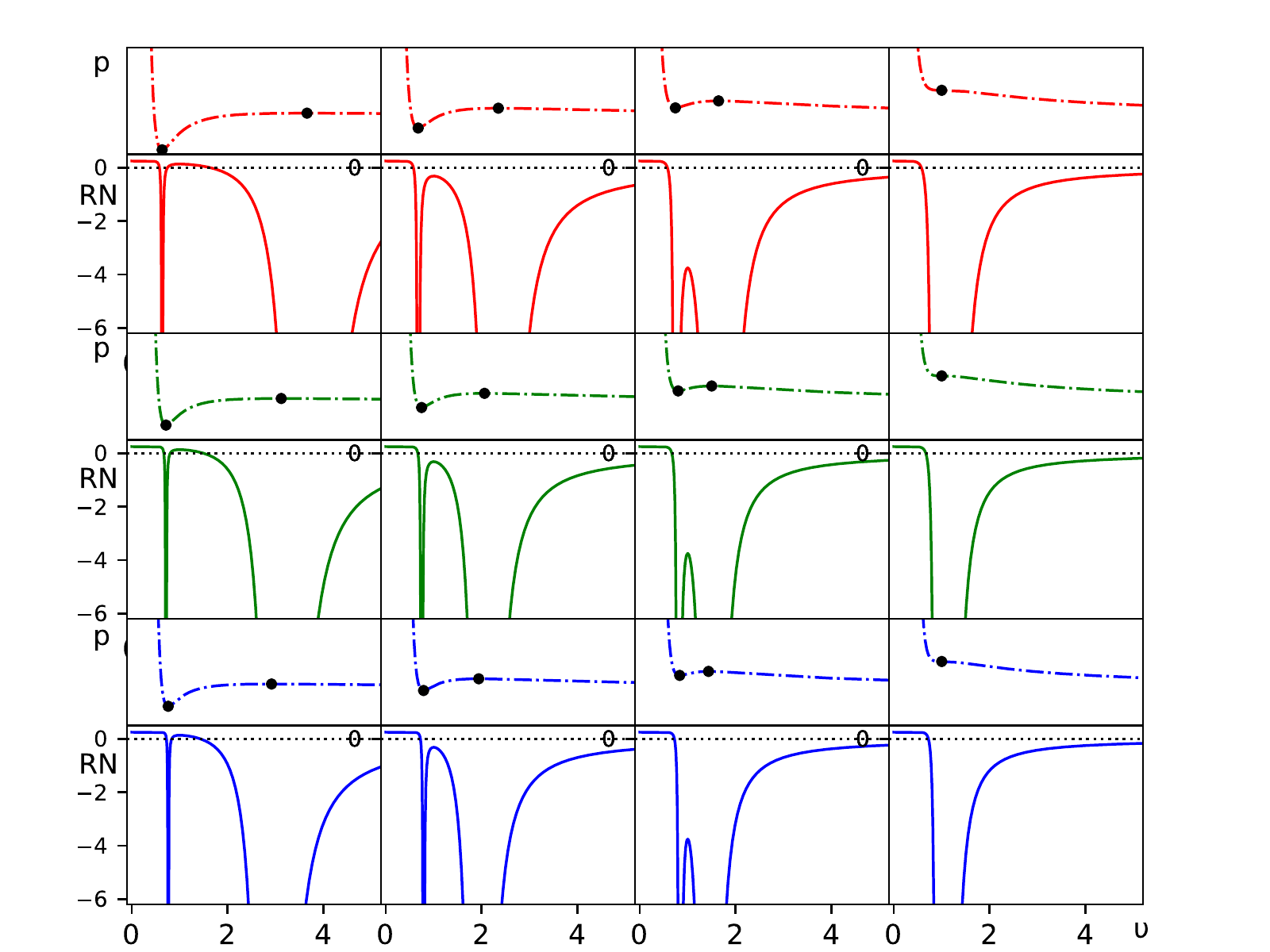}
\parbox[c]{15.0cm}{\footnotesize{\bf Fig~2.}  
${\rm R}_{\rm N}$ as a function of the $\nu$ with different temperatures. Each column from left to right represent $\tau=0.4$, $0.6$, $0.8$ and $1$. The top, middle, and bottom panels represent massive parameter $\lambda$ are $2$, $4$ and $6$.}
\label{fig2}
\end{center}

\par
On the other hand, considering the larger/smaller BH volume and Maxwell's equal area law, the reduced temperature is as follows
\begin{eqnarray}
\label{4-3}
\tau &=\frac{\lambda+1}{2\lambda\nu_{\rm l}}\Bigg(1+\frac{1}{\widetilde{x}}-\frac{1}{\widetilde{x}}\frac{\sum_{0}^{\lambda+1} \widetilde{x}^{n}}{\sum_{0}^{\lambda} {\rm d}_{\rm \lambda n}\widetilde{x}^{n}}\Bigg),\\
\label{4-4}
\tau &=\frac{\lambda+1}{2\lambda\nu_{\rm s}}\Bigg(1+\widetilde{x}-\frac{\sum_{0}^{\lambda+1} \widetilde{x}^{n}}{\sum_{0}^{\lambda} {\rm d}_{\rm \lambda n}\widetilde{x}^{n}}\Bigg),
\end{eqnarray}
and
\begin{eqnarray}
{\rm d}_{\rm \lambda 0}=1,\quad d_{\rm \lambda n}=d_{\rm \lambda n-1}+3(1-2\frac{n-1}{\lambda-1}).\nonumber
\end{eqnarray}

Assuming that $\tau \nu = k$, one can get
\begin{equation}
\label{4-5}
{\rm R}_{\rm N}=\frac{1}{2}-\frac{\lambda^2}{2\big(-\lambda+\frac{(\lambda+1)}{k}-\frac{\tau^{\lambda}}{k^{\lambda+1}}\big)^2}.
\end{equation}
At the critical point, $\widetilde{x}=1$ leads to $k=1$, corresponding to the normalized scalar curvature ${\rm R}_{\rm N}$ is
\begin{equation}
\label{4-6}
{\rm R}_{\rm Nc}=\frac{1}{2}-\frac{\lambda^2}{2(1-\tau^\lambda)^2}.
\end{equation}
Thus, by assigning a value to $k$ near 1, we can probe the characteristics of ${\rm R}_{\rm N}$ of the phase near the critical point. The left panel of Fig.3 shows that ${\rm R}_{\rm N}$ as a function of the $\tau$ with different $\lambda$. When $\tau<1$, the change rate of ${\rm R}_{\rm N}$ is more obviously affected by the parameter $\lambda$. The discrepancy of ${\rm R}_{\rm N}$ between two certain $k$ at constant $\tau$ becomes bigger when $\lambda$ increases. The right panel of Fig.3 shows that ${\rm R}_{\rm N}$ is a function of the $k$ with different $\lambda$. The discrepancy of ${\rm R}_{\rm N}$ between two certain $\tau$ at constant $k$ becomes bigger when $\lambda$ increases. Considering that the ${\rm R}_{\rm N}$ has a characteristic proportional to the dimension power of thermodynamic correlation length, $\lambda$ may be associated with thermodynamic correlation length, so that $\lambda$ showing an increase also increases the discrepancy of ${\rm R}_{\rm N}$.

\begin{center}
\includegraphics[width=7cm,height=5.5cm]{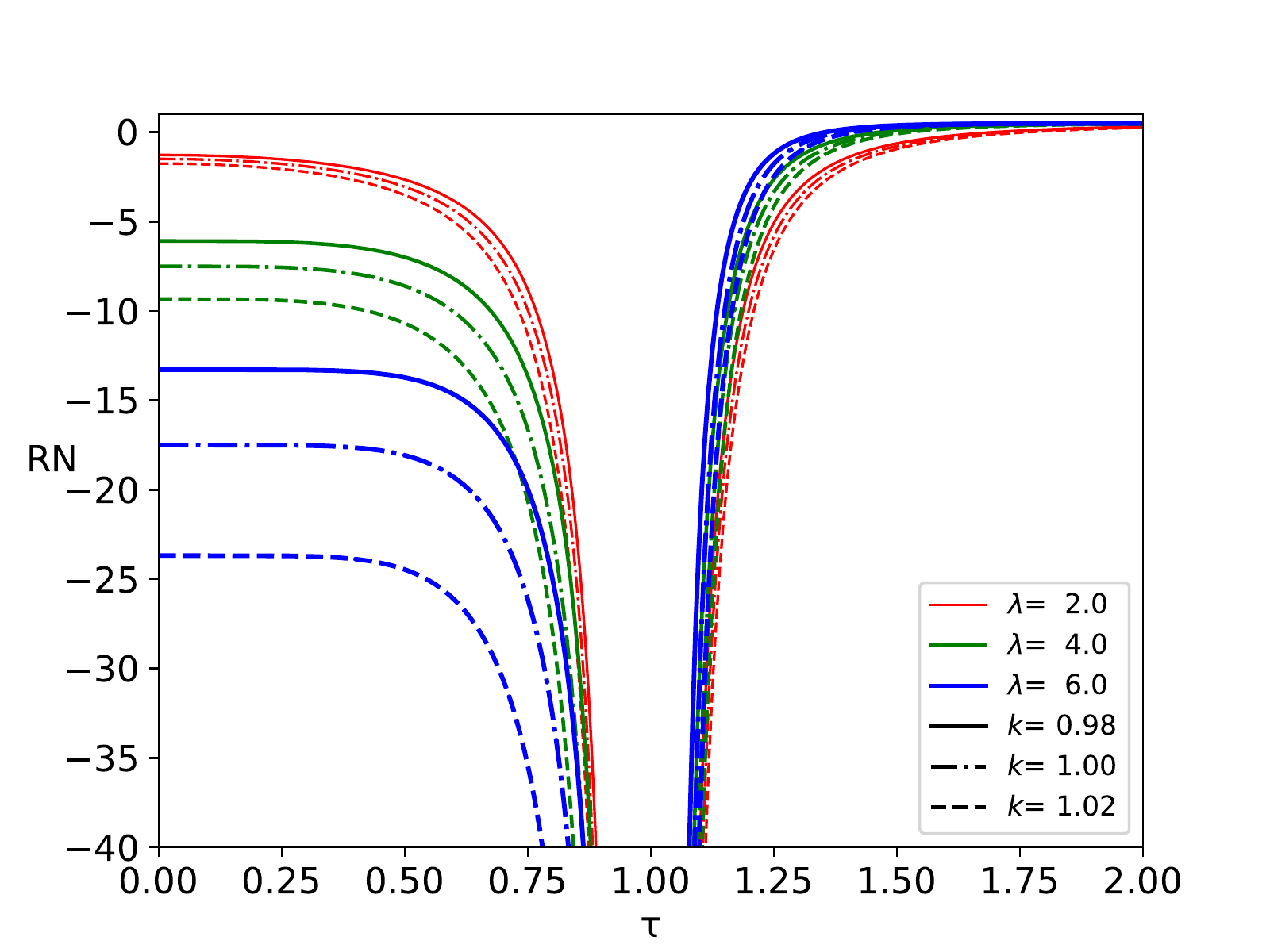}
\hspace{0.8cm}
\includegraphics[width=7cm,height=5.5cm]{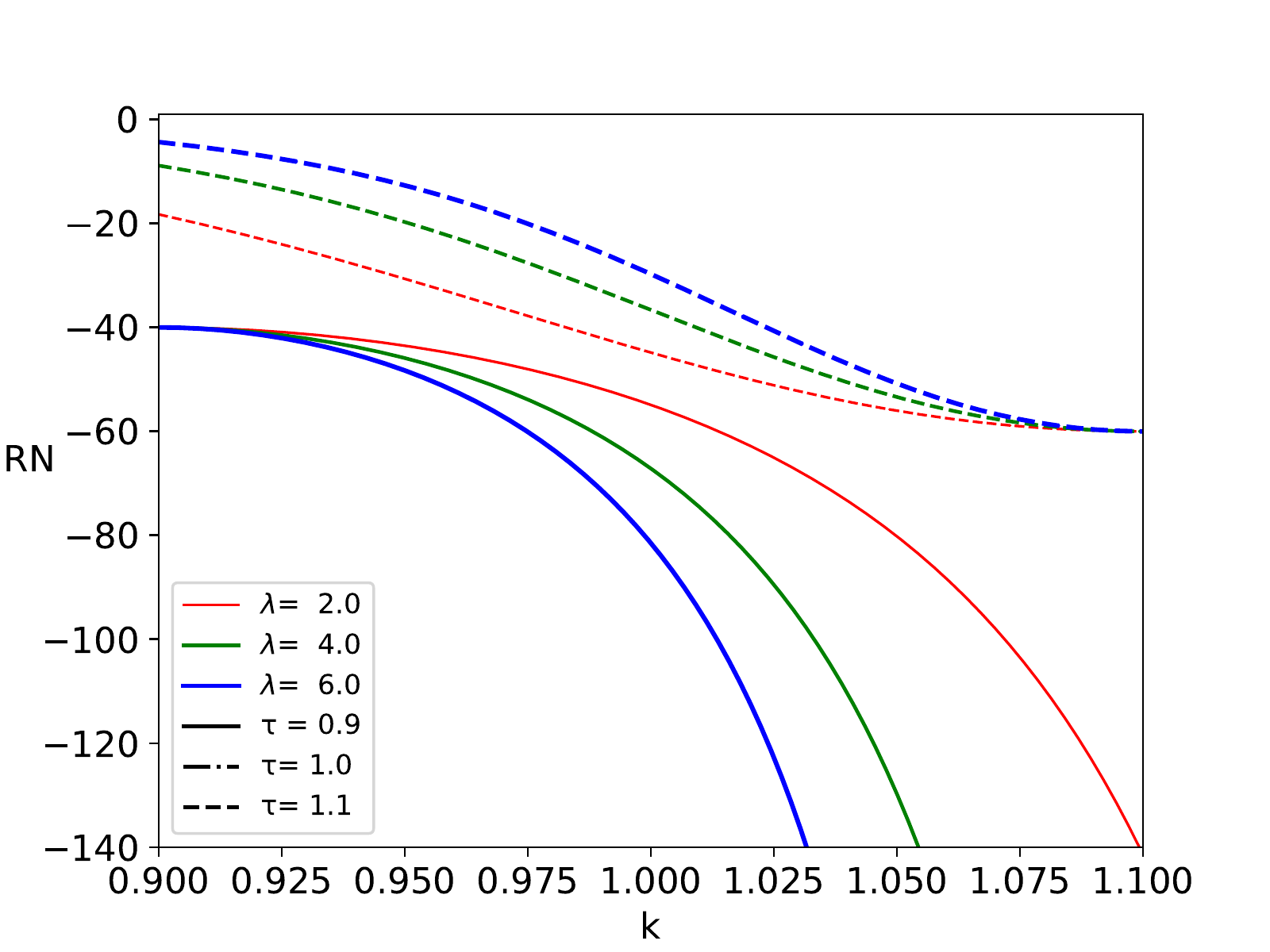}
\parbox[c]{15.0cm}{\footnotesize{\bf Fig~3.}  
Left: $R_{\rm N}$ as a function of the $\tau$ with different $\lambda$ at $k<1$, $k=1$ and $k>1$. Right: $R_{\rm N}$ as a function of the $k$ with different $\lambda$ at $\tau<1$, $\tau=1$ and $\tau>1$. The red, green and blue represent $\lambda$ parameters as $2,~4,~6$, respectively.}
\label{fig3}
\end{center}

\section{Conclusions and Discussion}
\label{sec:5}
\par
The phase transition grade and microstructure of the BH in massive gravity have been revealed in this analysis. We derived the equation of state and thermodynamic quantities of this BH in the extended phase space. By investigating isobaric heat capacity and Gibbs free energy, we found that the $C_{\rm P}$ function has transition and Gibbs function has swallowtail behavior indicating that the phase transition of this BH is a high-order phase transition. We obtained the critical universal constant and limited in $3/8< \varepsilon <1/2$.
\par
The more rational judgment of the phase transition grade in classical thermodynamics depends on the Ehrenfest equations. We verify the Ehrenfest equations at the phase transition critical point, showing the phase transition garde of BHs in massive gravity is second-order from macroscopic points of view. We obtain that the critical exponents of this BH are $(\alpha,~\beta,~\gamma,~\delta)=(0,~1/2,~1,~3)$, which is similar to the vdW system. Meanwhile, we found that the parameter $\lambda$ does not affect the BH phase transition grade. Based on the continuous phase transition theory, we analyzed the phase transition of BHs in massive gravity from the microcosmic point of view. The result suggests that the phase transition of BHs in massive gravity is a second-order phase transition from various perspectives, and it is still a vdW-like phase transition.
\par
In order to further understand the phase structure of BHs under the massive gravity framework, especially, the influence of parameter $\lambda$ of massive gravity theory on the phase near the critical point, we use Ruppeiner geometry to reveal the feature of the BH phases transition. We found that the two divergence points of normalized curvature ${\rm R}_{\rm N}$ tends to coincide when the BH temperature is equal to the $T_c$, and the sign of ${\rm R}_{\rm N}$ has changed from positive to negative. With the decrease of temperature, the distance between the two divergence points gradually becomes farther, and the transformation of the sign of ${\rm R}_{\rm N}$ is still once. When the temperature drops to a certain range, the distance between divergence points is getting larger and larger, but the sign of ${\rm R}_{\rm N}$ changed three times. It is found that $T > T_c$, $T = T_c$, and $T < T_c$ correspond to the cases of zero divergence point, one divergence point, and two divergence points of the ${\rm R}_{\rm N}$, respectively. Further, we investigate the effect of parameter $\lambda$ of massive gravity theory on the microstructure details. The different parameter $\lambda$ has a more obvious effect on ${\rm R}_{\rm N}$ on the phase transition side, and increasing $\lambda$ will increase the difference of ${\rm R}_{\rm N}$ between the two phases near the critical point. In the other words, parameter $\lambda$ can effectively affect the thermodynamic correlation length. By comparing with the results of the RN-AdS BH system in Ref. \cite{33,34}, it can be found that the main characteristics of the microstructure of phase transition caused by scalar charge are consistent with the charge-dependent phase transition. But for scalar charges, the microstructure has some additional details caused by parameter $\lambda$. Our results show that in the non-flat four-dimensional space-time, from the perspective of BH thermodynamics, it is difficult to distinguish between an RN BH and a Schwarzschild BH interacting with the mass gravitons, although there are differences in the microscopic details of their systems.
\par
In the near future, according to the data released in collaborations such as Laser-Interferometer Gravitational Wave-Observatory, Virgo Gravitational Wave- Interferometer (LIGO-VIRGO), and Einstein Horizon Telescope (EHT), the result of identifying mass gravitons near the black hole may become less difficult. The propagation speed of gravitational waves can provide information on the mass of graviton, and the dispersion relation of gravitational waves can characterize the Signs of Lorentz violating in massive gravity \cite{39,40,41}. For BH images taken by EHT, we may be able to get more intuitive results. Recently, it was proposed that the nonlinear characteristics of the electromagnetic field will affect the effective geometry of space-time, which may leave traces on BH images \cite{42,43}. The scalar charge caused by the graviton with mass should not have such a result.

\section*{Acknowledgments}
This work is supported by the National Natural Science Foundation of China (Grant No. 11903025).

\clearpage

\end{CJK}
\end{document}